# Lifetimes of Atmospheric Carbon Dioxide and Heat

PREPRINT

David A. Burton[a]

[a] *Burton Systems Software, Cary, NC*

*Corresponding author*: David A. Burton, dave@sealevel.info

ABSTRACT

Several different "lifetimes" are commonly claimed for atmospheric carbon dioxide, and confusion about them has led some people to erroneous conclusions about the sources and impacts of $CO_2$ emissions and changing atmospheric $CO_2$ levels. In this paper five different claimed $CO_2$ lifetimes are discussed, along with their impacts, the reasons that they differ, and their relevance to climate studies. Additionally, the lifetime of an addition of heat to the atmosphere is determined and discussed.

In the context of climatology, the most relevant of those lifetimes is the "adjustment time" of 45–65 years, which is determined from the observed roughly linear relationship between the net natural $CO_2$ removal rate and the atmospheric $CO_2$ level. That makes the effective half-life of added $CO_2$ 31–45 years.

## Introduction

The five commonly mentioned lifetimes of $CO_2$ in the atmosphere are (in order of increasing length):

1. The "residence time," of 3–5 years. The IPCC's AR6 report calls this "turnover time." It's the average time a $CO_2$ molecule remains in the atmosphere before being exchanged or removed. It is of little relevance for climate, but it has confused some authors into underestimating the effect of anthropogenic $CO_2$ emissions.

2. The observed $^{14}C$ radiocarbon "bomb spike lifetime" is an e-folding lifetime (decay time constant) of about twenty years (but sometimes mistakenly reported as 15–16 years). This has important ramifications for radiocarbon dating, but not for climate.

3. The measurement-derived "adjustment time" is an e-folding lifetime of 55 ± 10 years. This is the lifetime which determines the duration of effect for contemporary $CO_2$ emissions, and it is the primary relevant lifetime for climate studies.

4. Some authors cite a theoretical "long tail" atmospheric lifetime, often claimed to be hundreds or even thousands of years, based on the Bern model and similar models. The long tail lifetime would become important if $CO_2$ concentrations were falling, rather than rising.

5. Some studies assert that $CO_2$ emissions have a permanent, "cumulative" effect on temperatures, and therefore natural removals of $CO_2$ from the air can be ignored. In essence, they contend that the mere memory of $CO_2$ formerly in the atmosphere has the same impact on temperatures as the $CO_2$ which remains. That is unrealistic.

6. Most of the interest in $CO_2$ lifetimes is because of its effect on temperatures. But heat, itself, also has an atmospheric lifetime. The e-folding lifetime of heat (thermal energy) added to the atmosphere is about 52 days.

| **Which Lifetime** | **Duration** | **Significance** |
|---|---|---|
| 1. Turnover/residence time | About 4 years | Not significant for climate |
| 2. Bomb spike decay time constant | About 20 years | Important for radiocarbon dating, but not for climate |
| 3. Adjustment time | About 55 years | Determines duration of climate impact |
| 4. Long-tail lifetime | Hundreds or thousands of years | Would be significant if $CO_2$ levels were falling |
| 5. Cumulative effect | Effectively infinite | Unrealistic |
| 6. Lifetime of heat added to the atmosphere | About 52 days | Not significant for climate |

Table 1. Five reported "lifetimes" of atmospheric $CO_2$, and one for heat.

## 1. The short "residence time" or "turnover time"

The short residence time, which AR6 calls the "turnover time," is the average amount of time that molecules of $CO_2$ added to the atmosphere remain there. This is commonly estimated to be only 3–5 years (IPCC AR6 WG1 2021 Annex VII, p. 2237).

The reason it's so short is that, in addition to the processes which affect the other lifetimes, the turnover time is further shortened by processes which only temporarily remove carbon from the atmosphere.

For example, when $CO_2$ is absorbed by a puddle or raindrop, that event ends the residence time of those molecules in the air. When that puddle or raindrop evaporates, releasing the dissolved $CO_2$ back into the air, that's the start of a new residence time for the $CO_2$ molecules. But that round trip, from air to water and back, does not affect the amount of $CO_2$ in the air.

Similarly, when carbon is only briefly removed from the air into the terrestrial biosphere or ocean, and is then returned to the air, either immediately or perhaps after a growing season, that temporary removal "resets the timer" with regard to its residence time, but it does so while only transiently affecting the amount of $CO_2$ in the air.

Because those processes have no significant long-term effect on the amount of $CO_2$ in the air, they are not relevant when analyzing atmospheric $CO_2$ trends. So, the 3- to 5-year short residence time is irrelevant to projecting future atmospheric $CO_2$ trends or analysis of the causes of observed $CO_2$ trends.

Some authors (Harde 2017; Berry 2019; Stallinga 2023; Koutsoyiannis 2024) have concluded that the short residence time of anthropogenic $CO_2$ means that human $CO_2$ emissions have little effect on atmospheric $CO_2$ levels. They are mistaken (Engelbeen et al. 2024).

## 2. The $^{14}C$ radiocarbon "bomb spike" atmospheric lifetime

The atmospheric lifetime of "bomb spike" $^{14}C$ radiocarbon in the atmosphere is about 20 years. (It is sometimes erroneously estimated as 15–16 years.)

$^{14}C$ is usually reported as "$\Delta^{14}C$," which is defined as the molar fraction of atmospheric carbon which is (or was) in the form of the $^{14}C$ isotope, divided by a standard "Modern" value for that fraction (either $^{14}C/C = 1.170 \times 10^{-12}$ or $^{14}C/C = 1.176 \times 10^{-12}$), and reported as ‰ (parts-per-thousand) relative to that standard, either positive or negative (Schwartz et al. 2024).

The amount of $^{14}C$ radiocarbon in the atmosphere was nearly doubled by atmospheric testing of hydrogen bombs in the 1950s and early 1960s (i.e., $\Delta^{14}C$ briefly approached +1000‰ in the northern hemisphere). When the Limited Test Ban Treaty went into effect in 1963, the USA, USSR & UK ceased atmospheric tests of H-bombs, and the "bomb spike" soon peaked and began falling, as exchanges with the ocean and other carbon reservoirs replaced the atmospheric $CO_2$ containing elevated radiocarbon.

Additionally, by the late 1960s the atmospheric $CO_2$ level was rising by about 1 ppmv/year due to human emissions, mostly from fossil fuels. (Note: "ppmv" is shorthand for "µmol/mol of dry air.")

That so-called "fossil carbon" is depleted of $^{14}C$, because $^{14}C$ decays to $^{14}N$ with a half-life of 5700 ± 30 years, and fossil fuels are tens of thousands of times that old. So, adding fossil carbon to the atmosphere reduces the percentage (but not the total amount) of $^{14}C$ in $CO_2$ in the air. That is called the "Suess effect" (Keeling 1979). The Suess effect accounts for about one-fourth of the total observed decline in $\Delta^{14}C$ since the bomb spike.

As a result of the combined effects of those two processes, the $\Delta^{14}C$ bomb spike declined with an observed half-life of about 11 years.

At first blush, that would appear to make the atmospheric lifetime $11 / \ln(2) \approx 16$ years. However, increasing the amount of $CO_2$ in the atmosphere by burning fossil fuels reduces $\Delta^{14}C$ without actually removing $^{14}C$ from the air. It only reduces the fraction of carbon which is in the form of $^{14}C$.

To calculate the actual average atmospheric lifetime of $^{14}C$ added to the atmosphere by the bomb spike, we need to consider, instead, what the $\Delta^{14}C$ decay rate would have been, were it not for the Suess Effect reduction from fossil carbon emissions.

Graven et al. (2020) conducted simulations with a simple carbon cycle model, and plotted the calculated $^{14}CO_2$ decay with and without fossil fuel $CO_2$ supply, as shown in Fig. 1 (their Figure 4):

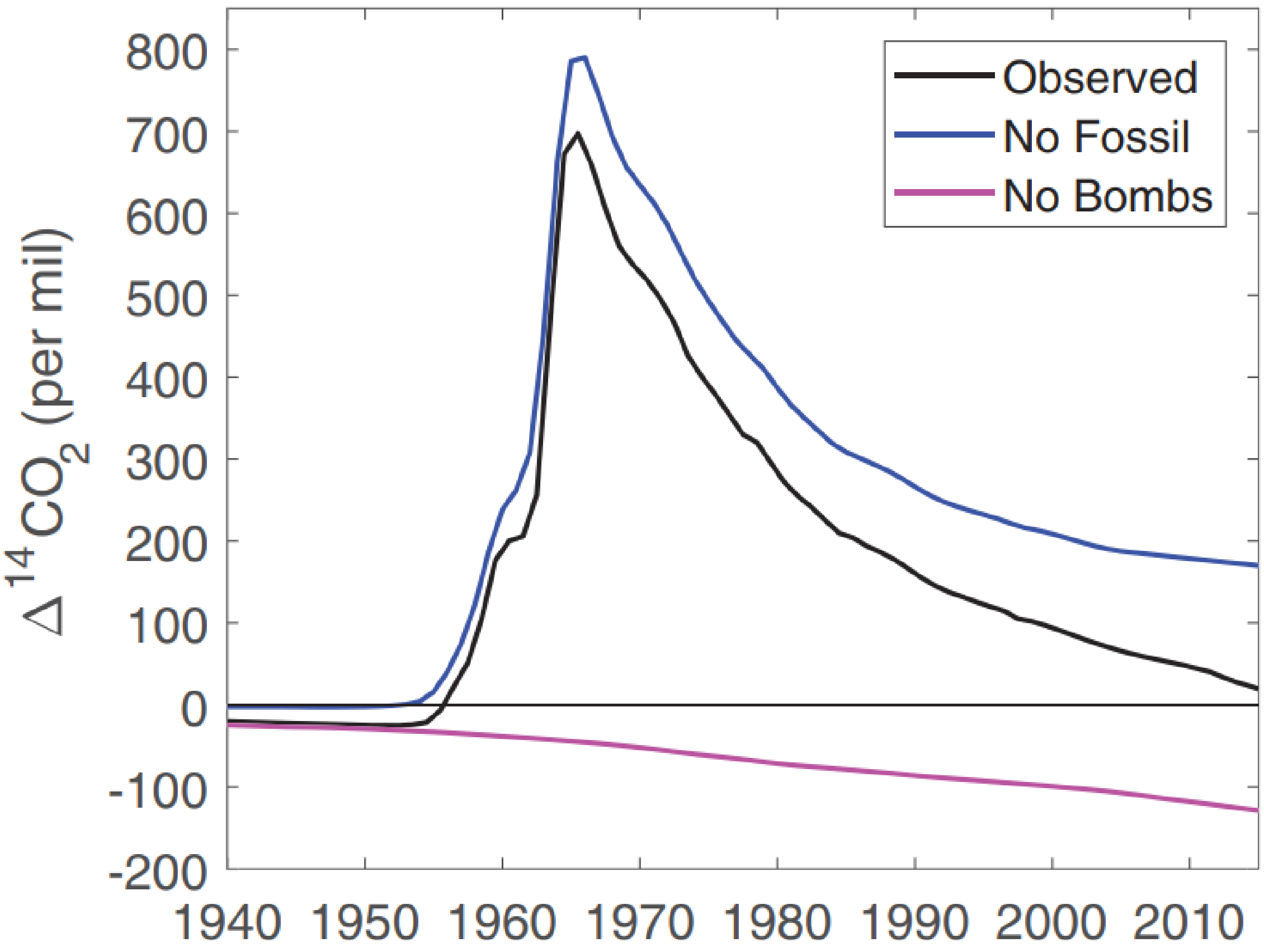

Fig. 1. $^{14}C$ radiocarbon bomb spike: "Observed $\Delta^{14}CO_2$ and simulated $\Delta^{14}CO_2$ for scenarios without nuclear weapons tests ('No Bombs') or without fossil fuel burning ('No Fossil')." Source: Graven et al. (2020).

H. Graven provided the data from their simulations, which were used to construct the log-linear plot in Fig. 2, and determine the e-folding lifetime of about 20 years.

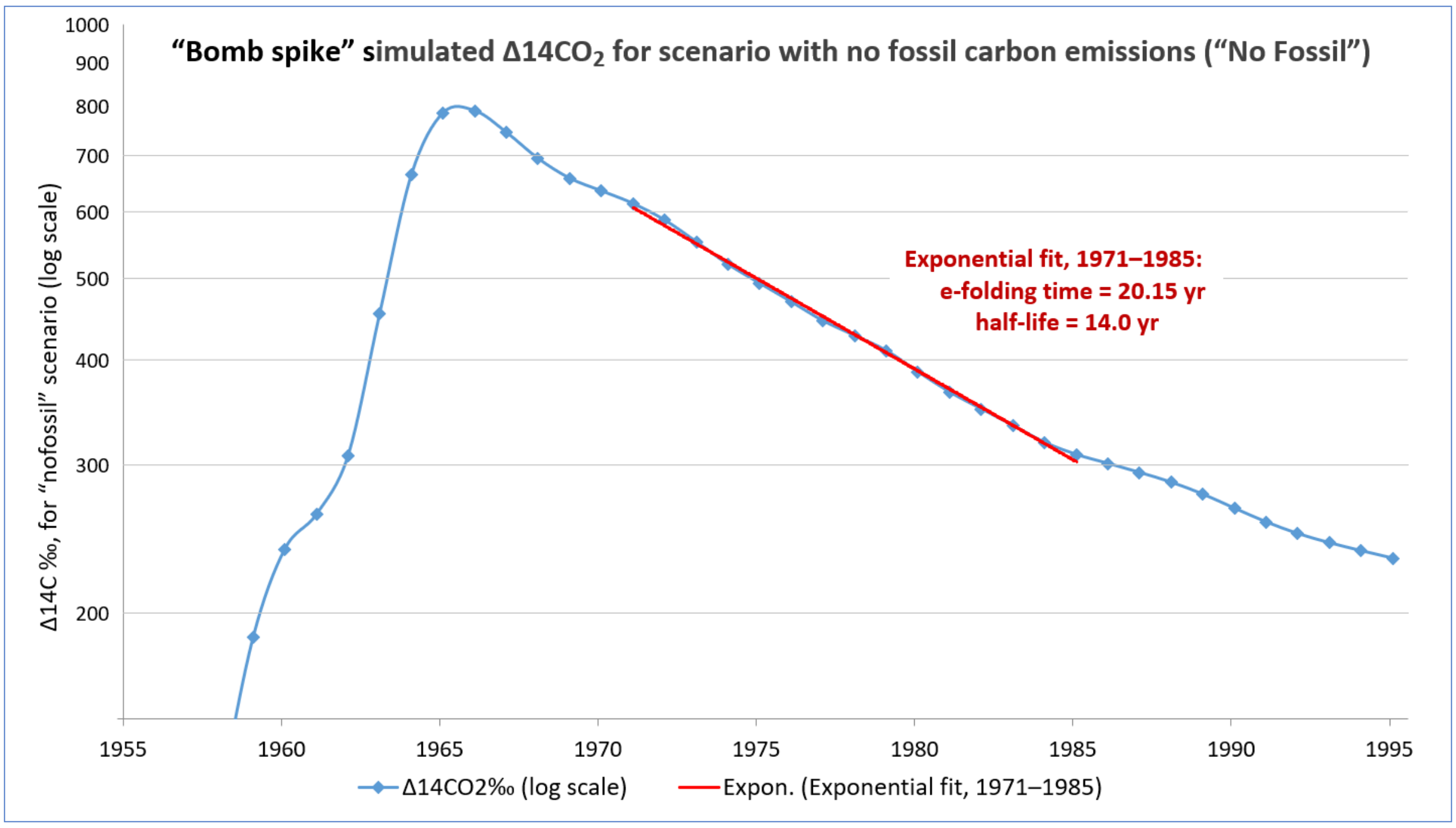

Fig. 2. Log-linear plot of "bomb spike" $\Delta^{14}C$, showing the approximately 20-year lifetime. Data is from the "no fossil" scenario depicted in Figure 4 of Graven et al. (2020). (The spreadsheet is included with the supplemental data.)

The reason for using the upper section of the decay curve for this determination is that it is the part of the measurement record for which the greatest portion of the observed trend is due to $^{14}C$ removals, which is what we are interested in, and for which the smallest portion of the observed trend is due to the Suess effect and the carbon cycle model.

The reason that the bomb spike radiocarbon atmospheric lifetime (20 years) is shorter than the adjustment time (55 ± 10 years) is that some of the processes which reduce the amount of $^{14}C$ in the atmosphere do so by exchanging it for carbon from other carbon reservoirs: mostly the ocean, the terrestrial biosphere and soils (Hua et al. 2022). (Physically, the sinks which remove atmospheric carbon and the exchanges which merely swap it for carbon from other reservoirs are mostly the same processes; the sinks represent the asymmetry between larger bidirectional fluxes.)

Exchanges of carbon between the atmosphere and other reservoirs reduce $\Delta^{14}C$ without reducing the total amount of carbon in the air. Thus they contributed to the relatively short 20-year atmospheric lifetime of "bomb spike" radiocarbon, but they do not reduce the amount of $CO_2$ in the atmosphere. So, the 20-year $^{14}C$ bomb spike lifetime is not the correct lifetime to use for projecting atmospheric $CO_2$ levels or for analyzing the causes of observed $CO_2$ level changes.

## 3. The measurement-derived adjustment time

The adjustment time can be thought of as the "effective residence time." It's the duration of the effect on $CO_2$ levels from adding $CO_2$ to the atmosphere.

The adjustment time is about 55 ± 10 years (half-life 38 ± 7 years). It is determined from observations, from the rate at which net natural sinks (mostly the ocean, terrestrial biosphere & soil) remove $CO_2$ from the atmosphere. Table 2 summarizes eleven adjustment time estimates; 55 ± 10 years encompasses the central values reported by all except one of them.

| **Study** | **Adjustment time, $\tau$** | **Half-life, $\tau \cdot \ln(2)$** |
|---|---|---|
| Moore and Braswell (1994) | 45 [27,71] years | 31 [19,49] |
| O'Neill et al. (1994) | 30–60 years | 21–42 years |
| Tahvonen et al. (1994) | 55 years | 38 years |
| IPCC SAR (1996) | 33–59 years | 23–41 years |
| Dietze (2001) | 55 years | 38 years |
| Schwartz (2019) | 54 ± 10 years | 30–44 years |
| Schwartz (2021) | 78–140 years | 54–97 years |
| Engelbeen (2022) | 50.5 years | 35 years |
| Hannesson (2023) | 56 years | 38 years |
| Spencer (2023) | 49.5 years | 34 years |
| this study | 58 years | 40 years |

Table 2. Eleven estimates of atmospheric $CO_2$ adjustment time.

The IPCC SAR (1996) reported it like this: *"Within 30 years about 40-60% of the CO2 currently released to the atmosphere is removed."*

If we call the fraction remaining in the atmosphere at 30 years F (0.6 to 0.4), that implies a half-life $T_{half}$ of $-30 \times \ln(2) / \ln(F)$ = 23 to 41 years, and an adjustment time of $T_{half} / \ln(2)$ = 33 to 59 years.

We now have another thirty years of data, so we can narrow that interval. We have good quality data for the rate at which fossil $CO_2$ is added to the atmosphere, which was about 4.9 ppmv in 2024, or 4.7 ppmv/year averaged over 2015–2024 (Friedlingstein et al. 2025). We also have excellent data since 1958 for the rate at which the atmospheric $CO_2$ level is rising, which is currently about 2.6 ppmv/year at Mauna Loa, averaged over a decade (Lan et al. 2026). The difference between those two rates is the net natural $CO_2$ removal rate: 4.7 – 2.6 = 2.1 ppmv/year.

Alternatively, similar calculations can be done which add to fossil carbon emissions estimates of "land use change emissions" ($CO_2$ released from clearing forests, draining swamps, etc.), and perhaps also subtract "cement carbonation sink" effects ($CO_2$ absorbed by concrete and mortar). That increases emissions and calculated removal rates by exactly the same amount (estimated by GCB 2025 as about 0.5 ppmv in 2024). However, estimates of land-use change effects and cement carbonation sinks are imprecise (Gasser et al. 2020; Dohner et al. 2022; Huang et al. 2023), so adding them increases the uncertainty intervals. I've chosen not to include them, which effectively treats them as diminishments of natural $CO_2$ removals, rather than as emissions.

Fig. 3 shows the net natural $CO_2$ removal rate plotted vs. the atmospheric $CO_2$ concentration, over the Mauna Loa measurement record (since 1958). It is evident that the $CO_2$ removal rate is an approximately linear function of the atmospheric $CO_2$ level.

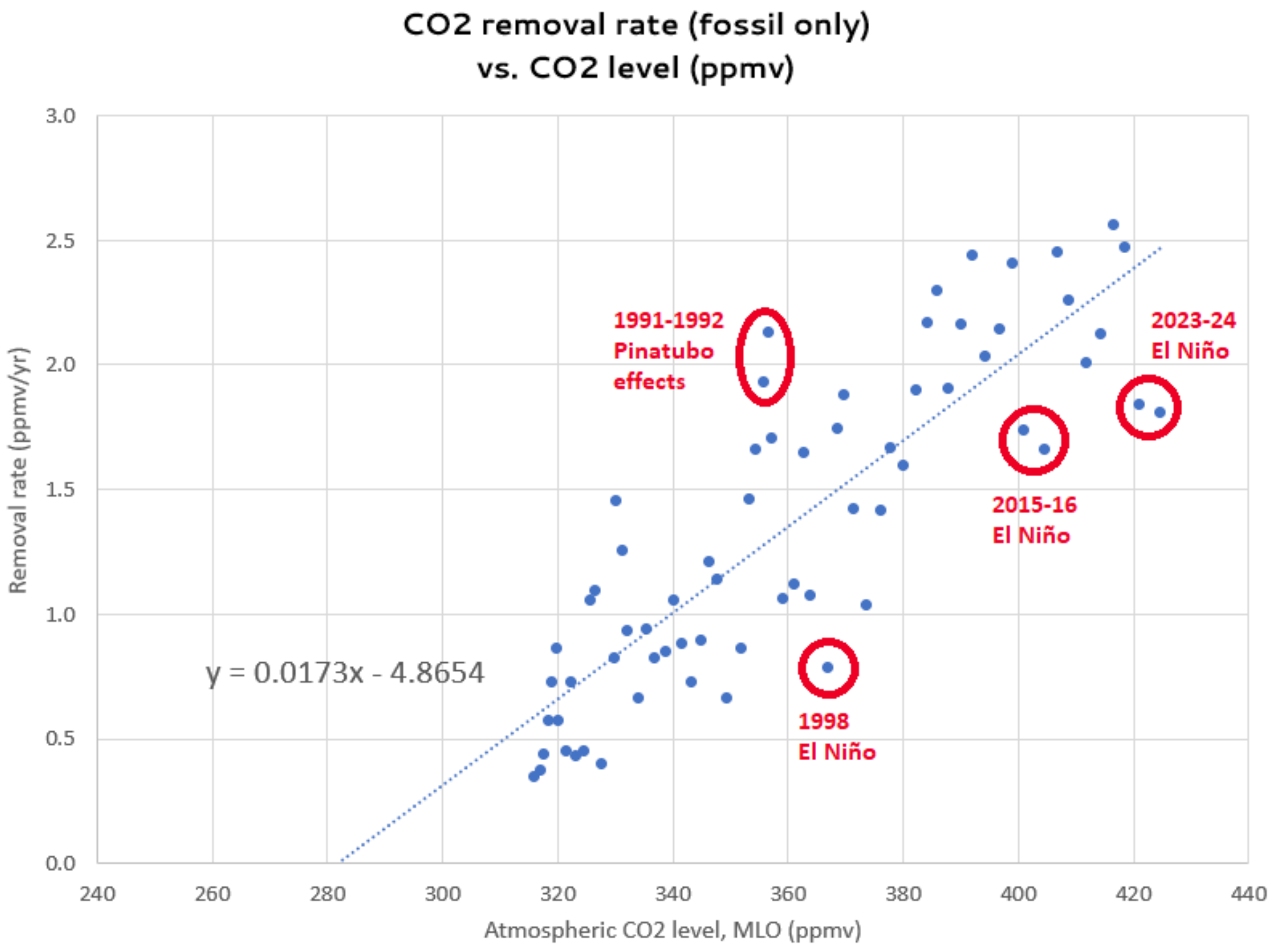

Fig. 3. Net natural $CO_2$ removal rate (fossil only) vs. Mauna Loa $CO_2$ level. Removal rates for each year are estimated as the difference between the previous year's and next year's average $CO_2$ levels divided by two, subtracted from the year's emissions. Removal rates are shown as ppmv/year; to convert to PgC/year multiply by 2.1294 PgC/ppmv. (The spreadsheet is included with the supplemental data.)

The x-intercept gives an estimate of the equilibrium level, which is about 282 ppmv. (Alternative calculations which include estimated land use change emissions yield somewhat different values.)

The slope is 1.73% per year. In other words, about 1.73% of the "excess" $CO_2$ above the approximately 282 ppmv equilibrium level is removed each year by natural sinks. That implies an adjustment time of about 1 / 0.0173 = 58 years.

That's probably a bit too high, because the data period ends with a strong El Niño, and hence a low removal rate. Spencer (2023) analyzed the relation between atmospheric $CO_2$ level and the rate of natural $CO_2$ removal, explicitly taking into account ENSO. He reported an equilibrium level (x-intercept) of 295 ppmv, and a slope (net natural $CO_2$ removal rate) of 2.02% per year (above that equilibrium level). A removal rate of 2% per year means the adjustment time is 1 / 0.02 = 50 years. A 50-year adjustment time makes the half-life of added $CO_2$ $50 \times \ln(2) \approx 35$ years.

The linearity of the relation between $CO_2$ level and removal rate, which is obvious in Fig. 3, is unsurprising, since the two major contributors, ocean uptake and terrestrial greening, are both approximately linear functions of atmospheric $CO_2$ level.

That may seem obvious for ocean uptake, since the $CO_2$ and dissolved inorganic carbon content at the ocean surface quickly equilibrates with the atmosphere, so elevated atmospheric $CO_2$ accelerates the migration of carbon from the surface water to the abyss via biological mechanisms (Rivero-Calle et al. 2015) and thermohaline circulation.

It is less obvious for greening. However, agronomists have extensively studied the responses to elevated $CO_2$ of a wide variety of plants, and those studies show that the average response of $C_3$ plants to elevated $CO_2$ is highly linear, until $CO_2$ levels exceed about 1000 ppmv, which is far higher than burning recoverable fossil fuels could ever raise outdoor $CO_2$ levels (Wang et al. 2017).

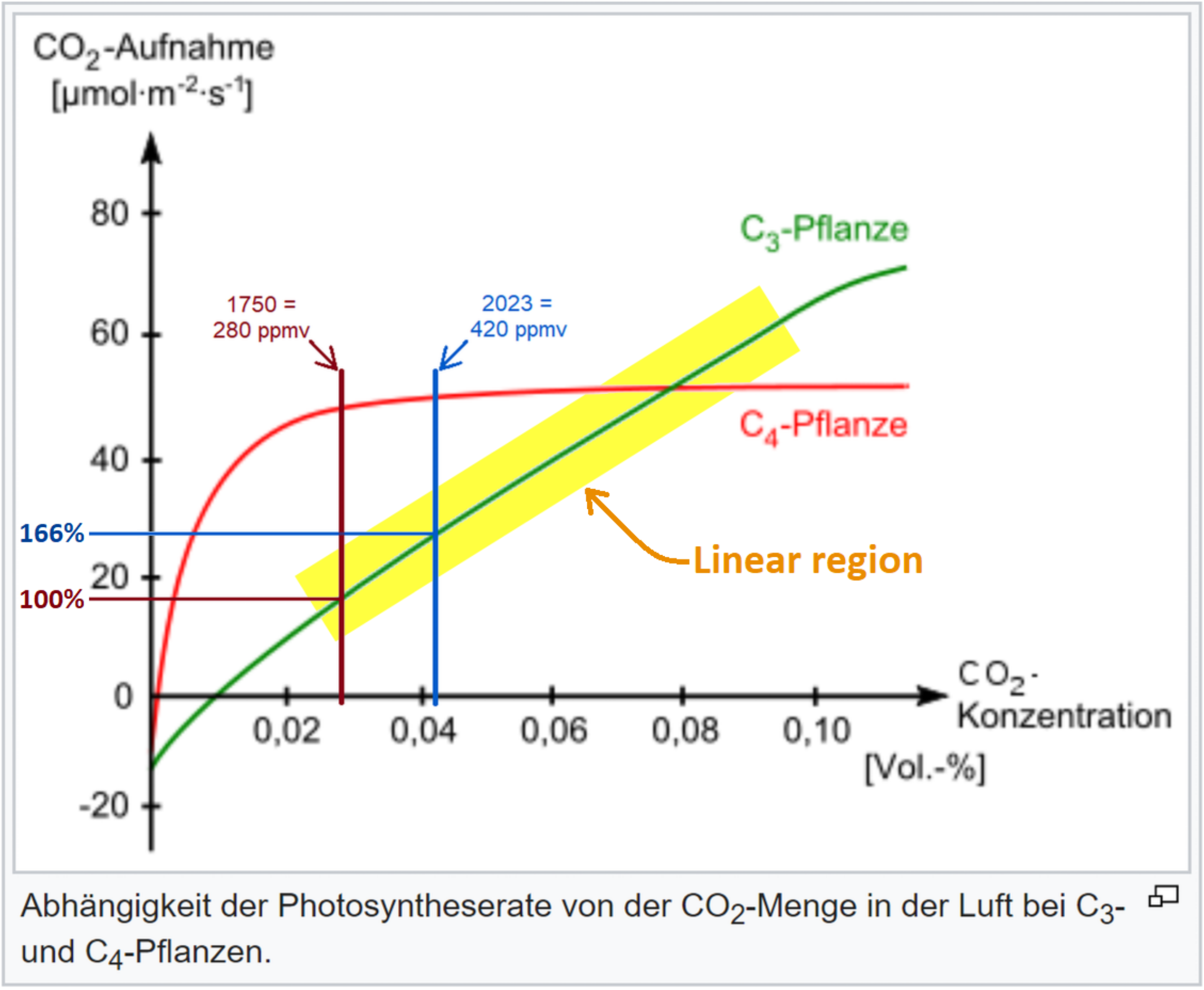

Fig. 4. Dependence of the rate of photosynthesis on the amount of $CO_2$ in the air in C3 and C4 plants. Adapted from Wikipedia reproduction of an illustration from Lüttge *et al.* (2005, 473); https://commons.wikimedia.org/wiki/File:Photosynthese_CO2-Konzentration.svg, CC BY-SA 3.0. Annotated by the author.

Fig. 4 is adapted from Lüttge *et al.* (2005), by way of Wikipedia. It shows that the average rate of $CO_2$ uptake by C3 plants is an approximately linear function of atmospheric $CO_2$ level until $CO_2$ levels exceed about 1000 ppmv. Note that trees and sphagnum moss, which sequester large amounts of carbon for relatively long periods of time, are C3 plants.

A 55-year adjustment time $\tau$ means that after $n$ years $\exp(-n/\tau) = \exp(-n/55)$ of the emissions remain. So, after 38 years (one half-life) about half of the increase in $CO_2$ level which resulted from this year's $CO_2$ emissions will be gone, and after 55 years about 63% of the increase due to this year's emissions will be gone. (The percentage of actual original "fossil" carbon atoms remaining in the air by then will be diminished even more, because of processes which exchange carbon in the air for carbon in other carbon reservoirs without changing the respective total amounts of carbon, but that's irrelevant for climate.)

It also means that for each 55 ppmv increase in the atmospheric $CO_2$ concentration, the net rate of natural $CO_2$ removals accelerates by about 1 ppmv/year.

That has important implications for climatology and public policy. For instance, since the $CO_2$ level is currently rising at about +2.6 ppmv/year, if the current emission rate were to continue indefinitely, the atmospheric $CO_2$ concentration would not continue to rise indefinitely. Instead, it would plateau at about $2.6 \times (55 \pm 10) = 143 \pm 26$ ppmv above its current level, because at that level the net rate of natural $CO_2$ removals would equal the anthropogenic emission rate.

It also means that if our fossil CO2 emissions were reduced from the current 5.0 ppmv/year to 2.4 ppmv/year, the CO2 level would no longer be rising (because emissions would be balanced by natural removals), and the AF would be zero. If they were reduced by more than that then the $CO_2$ level would be falling, and the AF would be negative.

If humans suddenly ceased emitting fossil $CO_2$ altogether then the $CO_2$ level in the atmosphere would decline at an initial rate of about 2.4 ppmv/year. (A simple simulation program is included in the supplemental material.)

The 55 ± 10 year adjustment time is the only time constant which matters for analyses of the cause of rising atmospheric $CO_2$ levels (Engelbeen et al. 2024), and for projections of future $CO_2$ levels, as long as $CO_2$ levels are rising rather than falling.

*a. A note about the "airborne fraction" (AF)*

The AF is commonly referred to as "the fraction of emissions that remain in the atmosphere" (IPCC AR6 WG1 2021 Annex VII, p. 2217; Bennedsen et al. 2024). That confuses many people, who mistakenly think it means that a constant fraction of emissions is somehow transported into the ocean and terrestrial biosphere, with the remainder (the AF) staying in the air indefinitely.

That is not what happens. There is no transport of a fixed fraction of anthropogenic emissions from chimney tops to other carbon reservoirs. All emissions go into the atmosphere.

The AF does not represent anything physical. It is just a computed ratio between the observed rate of $CO_2$ change in the atmosphere and the rate of anthropogenic emissions.

As Knorr (2009) observed, “Uptake of atmospheric $CO_2$ should react not to a change in emissions, but to a change in concentrations.” So, the fact that the AF ratio is currently about one-half is merely coincidental. In fact, it needn't even be positive.

*b. Application of adjustment time to climate modeling*

The adjustment time can be used to calculate estimates of the long-term effect of emission changes on atmospheric $CO_2$ level, radiative forcing (RF), and temperatures.

For example, suppose an emission source produces 1 Gt $CO_2$ per year. (That’s a little more than total emissions from Japan, or from the global civil aviation industry.) An adjustment time of 55 ± 10 years means that continual emissions of 1 Gt CO2/year will eventually increase the global atmospheric $CO_2$ level by 55 ± 10 Gt, which is about 55 / (7.8024 Gt $CO_2$/ppmv) = 7 ppmv. We could call that the “plateau change.”

Conversely, eliminating a 1 Gt per year $CO_2$ emission source would eventually reduce the atmospheric $CO_2$ level by about 7 ppmv.

By year 2100 (74 years = 1.35 e-folding lifetimes), it would have reached $1 - \exp(-n/\tau) = 1 - \exp(-74/55) = 74\%$ of the plateau change, or about 5 ppmv.

By year 2050 (24 years), it would have reached 35% of the plateau change, or 2.5 ppmv.

If we assume an eventual 600 ppmv $CO_2$ plateau level, an extra 7 ppmv would be $\log_2((600+7)/600) = 1.7\%$ of a doubling. If doubling $CO_2$ yields RF of 3.0 $W{\cdot}m^{-2}$ (Wijngaarden and Happer 2020) then 1.7% of a doubling would yield about 0.05 $W{\cdot}m^{-2}$, or if a doubling yields RF of 3.7 $W{\cdot}m^{-2}$ (Myhre et al. 1998) then 1.7% of that would be 0.06 $W{\cdot}m^{-2}$.

Similarly, if ECS climate sensitivity is 1.86 °C per doubling (R. W. Spencer and Christy 2024), 1.7% of that would be 0.03 °C, or if ECS climate sensitivity is 3 °C per doubling (IPCC AR6 WG1 2021, Sec. 7.5.5), 1.7% of that would be 0.05 °C.

## 4. The theoretical “long tail” lifetime

If $CO_2$ levels were falling, “browning” would eventually replace “greening” (Zhu et al. 2016), so the terrestrial biosphere would eventually become a source of $CO_2$, rather than a sink. So, models of the theoretical decay curve for atmospheric $CO_2$ in the hypothetical event that emissions suddenly cease typically show a "long tail" (Archer et al. 2009). That is, the decay curve would gradually diverge from a simple exponential decline, in that the second $CO_2$ half-life would be longer than the first, the third would be longer than the second, etc.

That is widely accepted. However, nature provides few clues for modeling the shape of the decay curve's long tail. So, the models and their projections are subject to considerable uncertainty, as well as being untestable.

Models like MAGICC (Meinshausen et al. 2011) or the Bern Model (Strassmann and Joos 2018) are widely used, probably because they're readily available, but there are many others, including Archer's CC_SED, the GENIE series, GEOCARB and GEOCYC, LTCM, MESMO, MPI0UW, and more (Archer et al. 2009). Unfortunately, there's no consensus between those models, so estimates for the "fatness" and "length" of the "long tail" vary wildly (Wang et al. 2017). Some authors claim that as much as 40% of a theoretical pulse of anthropogenic $CO_2$ will remain for more than 1000 years, or even that "the mean lifetime of fossil fuel $CO_2$ is about 30–35 kyr" (Archer 2005). So, integrating the modeled long tail yields very uncertain estimates of climate impact.

The long tail lifetime would mainly become relevant if $CO_2$ concentrations were falling, especially during the second and subsequent half-lives of the decay curve. However, second and subsequent $CO_2$ half-lives are largely irrelevant while $CO_2$ emissions exceed natural $CO_2$ removals and atmospheric $CO_2$ concentrations are rising, as is currently the case. Moreover, if $CO_2$ levels fall so low that second and subsequent half-lives become relevant, the consequent decline in agricultural yields will make it painfully clear that slowing the $CO_2$ decline is desirable, rather than harmful (O'Leary et al. 2015; Fitzgerald et al. 2016; Uddin et al. 2018; Taylor and Schlenker 2025).

So, the theoretical long-tail lifetime will be of little practical consequence for projecting climatological consequences of carbon emissions until $CO_2$ levels begin falling.

## 5. Cumulative effect

The "cumulative $CO_2$ emissions" framework is the theory that the warming effect of $CO_2$ is determined, not by the amount which remains in the atmosphere, but, rather, by the total amount which was ever emitted as a result of human activities. It is the basis for concepts like the "transient response to cumulative $CO_2$ emissions" (TCRE), and "remaining carbon budget" (RCB) (IPCC AR6 WG1 2021 Box TS.1, sections TS.1.3, TS.3.2.1, TS.3.3, 1.6.4, 5.5, Glossary, and SYN 3.3.1).

"Cumulative $CO_2$ emissions" does not mean the amount of $CO_2$ which has actually accumulated in the air. It's just a computed total of anthropogenic emissions to date.

The claim that cumulative emissions determine $CO_2$'s warming effect implies that the mere memory of $CO_2$ once in the atmosphere has as much warming effect as the $CO_2$ which remains. You could call that "$CO_2$ memory."

$CO_2$ memory is unphysical. The amount of $CO_2$ remaining in the atmosphere determines its warming effect (radiative forcing), not the total amount ever emitted.

The cumulative emissions framework ignores the natural processes which remove $CO_2$ from the air. Natural sinks are currently removing an average of more than 2 ppmv of $CO_2$ from the atmosphere per year. As shown in section 3, for each 55 ± 10 ppmv increase in the atmospheric $CO_2$ concentration the net rate of natural $CO_2$ removals accelerates by about 1 ppmv/year. That means the half-life of $CO_2$ added to the air is only 31–45 years. But the core assumption of the cumulative emissions framework is that those removals do not affect how much warming we get from $CO_2$.

TCRE and RCB, which are based on $CO_2$ memory, are themselves the basis for the "net zero" policy goal (rather than, say "net 50%"). TCRE and RCB are the topics of numerous peer-reviewed papers (Allen et al. 2009; Matthews et al. 2009; MacDougall and Friedlingstein 2015; Rogelj et al. 2019), and are unfortunately accepted by the IPCC. AR6 WG1 TS3.2.1 says, *"The transient climate response to cumulative $CO_2$ emissions (TCRE) is the ratio between globally averaged surface temperature increase and cumulative $CO_2$ emissions… This Report reaffirms with high confidence the finding of AR5 that there is a near-linear relationship between cumulative $CO_2$ emissions and the increase in global average temperature caused by $CO_2$ over the course of this century… in the 1.0°C–2.3°C per 1000 PgC range…"*

In other words, the correlation between two steadily rising quantities gives them "high confidence" that one causes the other. But, physically, only $CO_2$ still in the atmosphere can contribute to the radiative forcing which warms the Earth, so any apparent proportionality of warming to cumulative emissions cannot be causal.

A justification sometimes given for the cumulative emissions framework is that a fixed "airborne fraction" (AF) of anthropogenic $CO_2$ emissions remain in the atmosphere (IPCC AR6 WG1 2021, 2223 Glossary). But, as discussed in section 3a, the fixed AF is a fallacy.

The migration of $CO_2$ from the air into other carbon reservoirs is from the bulk atmosphere, not from a fixed fraction of emissions. The net rate of those removal processes depends mostly on the atmospheric $CO_2$ concentration, not on the emission rate. The fact that that rate is currently about half the rate of anthropogenic emissions is coincidental: a consequence of the fact that anthropogenic emissions have been increasing exponentially (Stokes 2015).

With one minor exception, the rates of removal processes are governed mainly by the atmospheric $CO_2$ level, somewhat modulated by factors like ENSO and temperatures.

The minor exception is the increase in carbon content of the upper ocean, a removal process which is governed by the rate of change of atmospheric $CO_2$, rather than its level. The dissolved inorganic carbon (DIC) content of the upper ocean closely tracks the atmospheric $CO_2$ level, so the rate at which upper-ocean DIC increases is proportional to the rate of change of atmospheric $CO_2$. Since the upper ocean holds roughly as much carbon as the atmosphere does, one might expect this process to be a major sink for atmospheric $CO_2$. But it is not, because carbonate buffering, quantified by the Revelle factor, reduces the carbon uptake by roughly a factor of ten (Revelle and Suess 1957).

There are no significant $CO_2$ removal processes which are governed by the rate of anthropogenic emissions.

So, if emissions were decreased to substantially less than the net natural $CO_2$ removal rate, in the short term the removal rate would hardly change. That means the removal rate would exceed the emission rate, so the atmospheric $CO_2$ level (and its warming effect) would be decreasing, and the "airborne fraction" would be negative. Despite cumulative emissions continuing to rise, and $CO_2$ emissions not declining to "net zero," the warming effect (radiative forcing) of atmospheric $CO_2$ would be decreasing.

The "$CO_2$ memory" basis of the cumulative $CO_2$ emission framework is uncomfortably reminiscent of "water memory," which is the pseudoscientific basis of homeopathy.

Homeopathic medicine is a field clothed with all the trappings of real science, including academic journals, peer-reviewed papers, widespread legal recognition, professional societies and organizations, and even college programs & degrees. The journal *Homeopathy* (doi:

10.1055/s-00034926), which publishes papers on homeopathic medicine, was founded in 1911. It has long been published by reputable publishers (until 2018 by Elsevier; now by Thieme). Yet the field has no scientific validity.

Homeopathy is based on the unphysical concept of “water memory.” Homeopaths contend that if a therapeutic agent is dissolved in water, and then the water is repeatedly diluted, so many times that not even a single molecule of the agent remains, the preparation can still have a potent therapeutic effect, because the water “remembers” what used to be dissolved in it—which is physically impossible.

Neither is the warming effect of $CO_2$ proportional to “cumulative” $CO_2$ emissions. The warming effect is proportional to the logarithm of the amount of $CO_2$ remaining in the atmosphere, and the amount of $CO_2$ remaining in the atmosphere is primarily governed by the adjustment time of 55 ± 10 years.

## 6. Lifetime of heat added to the atmosphere

Just as gases like $CO_2$ have finite lifetimes when added to the atmosphere, so too does heat.

Direct additions to the atmosphere of “anthropogenic heat” (thermal energy from burning fossil fuels and running nuclear power plants), are small relative to the radiative forcings from anthropogenic GHG emissions. Mankind currently produces about 600 Exajoules ($6.0 \times 10^{20}$ J) of energy per year, of which about 99% ultimately ends up as thermal energy. (About 6 EJ = 1% of it becomes stored chemical energy, mostly through aluminum smelting.) Dividing 600 EJ by the $5.10072 \times 10^{14}$ $m^2$ surface area of the Earth, and by the number of seconds in a year, yields an average of only 0.0373 $W{\cdot}m^{-2}$, compared to a net anthropogenic radiative forcing of between 2 and 3 $W{\cdot}m^{-2}$. So direct anthropogenic heat emissions have a negligible effect on the Earth’s temperature, as a whole (though it can be significant locally).

Nevertheless, it is interesting to consider the lifetime of thermal energy added to the atmosphere. Imagine a scenario in which an extraordinary event somehow suddenly raised the temperature of all the air in the Earth's atmosphere by 1°C. How long would it take for half of that increase to decay (i.e., for the air to cool by an average of 0.5°C)?

First, let's calculate how much energy is required to heat all the air in the Earth's atmosphere by 1°C.

The Earth's atmosphere is estimated to have a mass of about $5.15 \times 10^{18}$ kg (Trenberth and Smith 2005). Air has a heat capacity of 1.005 $kJ{\cdot}kg^{-1}{\cdot}K^{-1}$ at constant pressure, so to heat that much air by 1°C would require $(5.15 \times 10^{18}$ kg$) \times (1005\ J{\cdot}kg^{-1}) = 5.18 \times 10^{21}$ J.

However, assuming constant relative humidity, a 1°C warmer atmosphere would also contain 6% to 7% more water vapor, and the latent heat required to add that much water vapor increases the heat capacity of the air by about 40%. That would increase the effective heat capacity of the atmosphere from $5.18 \times 10^{21}\ J{\cdot}K^{-1}$ to about $(7.3 \pm 0.3) \times 10^{21}\ J{\cdot}K^{-1}$.

Typical estimates are that between 3.0 and 3.7 $W{\cdot}m^{-2}$ is needed to "raise temperatures" (which really means to *keep* temperatures raised!) by 1°C. The higher figure is calculated directly from the Stefan-Boltzmann relation, but feedbacks lower it slightly (Cronin and Dutta 2023). It is estimated in AR6 §7.4.2.1 to be 3.0 to 3.4 $W{\cdot}m^{-2}$ (IPCC AR6 WG1 2021).

Accepting that estimate, the amount of energy required to keep the atmosphere's temperature elevated by an average of 1°C is $(3.2 \pm 0.2)\ W{\cdot}m^{-2}$ × the surface area of the Earth × the number of seconds in a year = $(3.2 \pm 0.2)\ W{\cdot}m^{-2} \times (5.10072 \times 10^{14}\ m^2) \times$ 31,556,952 $s{\cdot}yr^{-1} = (5.151 \pm 0.322) \times 10^{22}$ J per year.

Dividing the heat capacity of the atmosphere by the energy per year required to maintain a 1°C temperature elevation yields the lifetime:

$$\frac{(7.3 \pm 0.3) \times 10^{21}\ J{\cdot}K^{-1}}{(5.151 \pm 0.322) \times 10^{22}\ J{\cdot}K^{-1}{\cdot}yr^{-1}} = 0.143 \pm 0.015 \text{ year} = 52 \pm 5 \text{ days}$$

The half-life of a temperature perturbation is ln(2) times the lifetime, which is about 36 days. Unsurprisingly, that is similar to the observed seasonal lag between solstices and the hottest or coldest days of summer or winter, respectively: roughly one month.

52 ± 5 days is also similar to the 58-day lifetime calculated by Osácar (2020) using a different approach.

## 7. Conclusions

In the context of climatology, the primary relevant "lifetime" of $CO_2$ emitted into the atmosphere is the "adjustment time," which is about 55 ± 10 years. That corresponds to a half-life of about 31–45 years. It is a consequence of the observed fact that the net rate of natural $CO_2$ removals from the atmosphere accelerates linearly by about 1 ppmv/year for each 55 ppmv rise in atmospheric $CO_2$ concentration.

Several other "lifetimes" of atmospheric $CO_2$ are definable, but using them for climate studies is a mistake, as is basing climate predictions on "cumulative emissions" of $CO_2$. However, the theoretically predicted "long tail" lifetime would become relevant if atmospheric $CO_2$ concentrations were falling, rather than rising.

The lifetime of heat added to the atmosphere is less than 2 months, so direct anthropogenic heat production, for example by thermal power plants, has a negligible effect on the Earth's average temperature.

*Acknowledgments.*

I am grateful to Heather Graven for sharing her radiocarbon decay simulation data; to Ferdinand Engelbeen, Renee Hannon, Greg Wrightstone, and Rögnvaldur Hannesson for helpful review and suggestions; and to Ferdinand Engelbeen, Renee Hannon, and Will Happer for teaching me many things.

I received no financial support for this work, and declare no conflicts of interest.

*Data Availability Statement.*

I am committed to full, open and timely access to code and data. All code, spreadsheets, and other data used and generated in this work are archived in the Zenodo repository at https://doi.org/10.5281/zenodo.23318629

Subsequent development versions will be available at https://sealevel.info/CO2_lifetimes/

## REFERENCES

Allen, Myles R., David J. Frame, Chris Huntingford, et al. 2009. "Warming Caused by Cumulative Carbon Emissions towards the Trillionth Tonne." *Nature* 458 (7242): 1163–66. https://doi.org/10.1038/nature08019.

Archer, David. 2005. "Fate of Fossil Fuel CO2 in Geologic Time." *Journal of Geophysical Research: Oceans* 110 (C9). https://doi.org/10.1029/2004JC002625.

Archer, David, Michael Eby, Victor Brovkin, et al. 2009. "Atmospheric Lifetime of Fossil Fuel Carbon Dioxide." *Annual Review of Earth and Planetary Sciences* 37 (Volume 37, 2009): 117–34. https://doi.org/10.1146/annurev.earth.031208.100206.

Bennedsen, Mikkel, Eric Hillebrand, and Siem Jan Koopman. 2024. "A Regression-Based Approach to the CO2 Airborne Fraction." *Nature Communications* 15 (1): 8507. https://doi.org/10.1038/s41467-024-52728-1.

Berry, Edwin X. 2019. "Human CO2 Emissions Have Little Effect on Atmospheric CO2." *International Journal of Atmospheric and Oceanic Sciences* 3 (1): 13–26. https://doi.org/10.11648/j.ijaos.20190301.13.

Cronin, Timothy W., and Ishir Dutta. 2023. "How Well Do We Understand the Planck Feedback?" *Journal of Advances in Modeling Earth Systems* 15 (7): e2023MS003729. https://doi.org/10.1029/2023MS003729.

Dietze, Peter. 2001. "Carbon Model Calculations." *Still Waiting for Greenhouse, by John L. Daly*. http://www.john-daly.com/dietze/cmodcalc.htm.

Dohner, Julia L., Benjamin Birner, Armin Schwartzman, Julia Pongratz, and Ralph F. Keeling. 2022. "Using the Atmospheric CO2 Growth Rate to Constrain the CO2 Flux from Land Use and Land Cover Change since 1900." *Global Change Biology* 28 (24): 7327–39. https://doi.org/10.1111/gcb.16396.

Engelbeen, Ferdinand. 2022. "Origin of the Recent CO2 Increase in the Atmosphere." http://www.ferdinand-engelbeen.be/klimaat/co2_origin.html.

Engelbeen, Ferdinand, Renee Hannon, and David Burton. 2024. *The Human Contribution to Atmospheric Carbon Dioxide*. CO2 Coalition. https://doi.org/10.31219/osf.io/het6n.

Fitzgerald, Glenn J., Michael Tausz, Garry O'Leary, et al. 2016. "Elevated Atmospheric [CO2] Can Dramatically Increase Wheat Yields in Semi-Arid Environments and Buffer against Heat Waves." *Global Change Biology* 22 (6): 2269–84. https://doi.org/10.1111/gcb.13263.

Friedlingstein, Pierre, Michael O'Sullivan, Matthew W. Jones, et al. 2025. "Global Carbon Budget 2025." *Earth System Science Data* 18 (5): 3211–88. https://doi.org/10.5194/essd-18-3211-2026.

Gasser, Thomas, Léa Crepin, Yann Quilcaille, Richard A. Houghton, Philippe Ciais, and Michael Obersteiner. 2020. "Historical $CO_2$ Emissions from Land Use and Land Cover Change and Their Uncertainty." *Biogeosciences* 17 (15): 4075–101. https://doi.org/10.5194/bg-17-4075-2020.

Graven, Heather, Ralph F. Keeling, and Joeri Rogelj. 2020. "Changes to Carbon Isotopes in Atmospheric CO2 Over the Industrial Era and Into the Future." *Global Biogeochemical Cycles* 34 (11): e2019GB006170. https://doi.org/10.1029/2019GB006170.

Hannesson, Rögnvaldur. 2023. The Airborne Fraction of Anthropogenic CO2 Emissions: A Mistaken Concept? ResearchGate. https://www.researchgate.net/publication/371220923_The_airborne_fraction_of_anthropogenic_CO2_emissions_A_mistaken_concept.

Harde, Hermann. 2017. "Scrutinizing the Carbon Cycle and CO2 Residence Time in the Atmosphere." *Global and Planetary Change* 152 (May): 19–26. https://doi.org/10.1016/j.gloplacha.2017.02.009.

Hua, Quan, Jocelyn C. Turnbull, Guaciara M. Santos, et al. 2022. "Atmospheric Radiocarbon for the Period 1950–2019." *Radiocarbon* 64 (4): 723–45. https://doi.org/10.1017/RDC.2021.95.

Huang, Zi, Jiaoyue Wang, Longfei Bing, et al. 2023. "Global Carbon Uptake of Cement Carbonation Accounts 1930–2021." *Earth System Science Data* 15 (11): 4947–58. https://doi.org/10.5194/essd-15-4947-2023.

IPCC AR6 WG1. 2021. *Climate Change 2021 – The Physical Science Basis: Working Group I Contribution to the Sixth Assessment Report of the Intergovernmental Panel on Climate Change*. 1st ed. Cambridge University Press. https://doi.org/10.1017/9781009157896.

IPCC SAR WG1. 1996. *Climate Change 1995 – The Science of Climate Change: Contribution of Working Group I to the Second Assessment Report of the Intergovernmental Panel on Climate Change. Technical Summary.* Cambridge University Press. https://www.ipcc.ch/report/ar2/wg1/.

Keeling, Charles D. 1979. "The Suess Effect: 13Carbon-14Carbon Interrelations." *Environment International* 2 (4): 229–300. https://doi.org/10.1016/0160-4120(79)90005-9.

Knorr, Wolfgang. 2009. "Is the Airborne Fraction of Anthropogenic CO2 Emissions Increasing?" Sec. L21710. *Geophysical Research Letters* 36 (21). https://doi.org/10.1029/2009GL040613.

Koutsoyiannis, Demetris. 2024. "Stochastic Assessment of Temperature–CO2 Causal Relationship in Climate from the Phanerozoic through Modern Times." *Mathematical Biosciences and Engineering* 21 (7): 6560–602. https://doi.org/10.3934/mbe.2024287.

Lan, X., P. Tans, and K. W. Thoning. 2026. "Atmospheric Carbon Dioxide Dry Air Mole Fractions from the NOAA GML Carbon Cycle Greenhouse Gases Group." NOAA Global Monitoring Laboratory. https://doi.org/10.15138/9N0H-ZH07.

Lüttge, Ulrich, Manfred Kluge, and Gabriela Bauer. 2005. *Botanik*. 5th ed. Wiley-Blackwell.

MacDougall, Andrew H., and Pierre Friedlingstein. 2015. "The Origin and Limits of the Near Proportionality between Climate Warming and Cumulative CO2 Emissions." Journal of Climate. *Journal of Climate* 28 (10): 4217–30. https://doi.org/10.1175/JCLI-D-14-00036.1.

Matthews, H. Damon, Nathan P. Gillett, Peter A. Stott, and Kirsten Zickfeld. 2009. "The Proportionality of Global Warming to Cumulative Carbon Emissions." *Nature* 459 (7248): 829–32. https://doi.org/10.1038/nature08047.

Meinshausen, M., S. C. B. Raper, and T. M. L. Wigley. 2011. "Emulating Coupled Atmosphere-Ocean and Carbon Cycle Models with a Simpler Model, MAGICC6 – Part 1:

Model Description and Calibration." *Atmospheric Chemistry and Physics* 11 (4): 1417–56. https://doi.org/10.5194/acp-11-1417-2011.

Moore III, Berrien, and B. H. Braswell. 1994. "The Lifetime of Excess Atmospheric Carbon Dioxide." *Global Biogeochemical Cycles* 8 (1): 23–38. https://doi.org/10.1029/93GB03392.

Myhre, Gunnar, Eleanor J. Highwood, Keith P. Shine, and Frode Stordal. 1998. "New Estimates of Radiative Forcing Due to Well Mixed Greenhouse Gases." *Geophysical Research Letters* 25 (14): 2715–18. https://doi.org/10.1029/98GL01908.

O'Leary, Garry J., Brendan Christy, James Nuttall, et al. 2015. "Response of Wheat Growth, Grain Yield and Water Use to Elevated CO2 under a Free-Air CO2 Enrichment (FACE) Experiment and Modelling in a Semi-Arid Environment." *Global Change Biology* 21 (7): 2670–86. https://doi.org/10.1111/gcb.12830.

O'Neill, Brian C., Stuart R. Gaffin, Francesco N. Tubiello, and Michael Oppenheimer. 1994. "Reservoir Timescales for Anthropogenic CO2 in the Atmosphere." *Tellus B: Chemical and Physical Meteorology* 46 (5). https://doi.org/10.3402/tellusb.v46i5.15812.

Osácar, Carlos, Manuel Membrado, and Amalio Fernández-Pacheco. 2020. "Brief Communication: Residence Time of Energy in the Atmosphere." *Nonlinear Processes in Geophysics* 27 (2): 235–37. https://doi.org/10.5194/npg-27-235-2020.

Revelle, Roger, and Hans E. Suess. 1957. "Carbon Dioxide Exchange Between Atmosphere and Ocean and the Question of an Increase of Atmospheric CO2 during the Past Decades." *Tellus* 9 (1): 18–27. https://doi.org/10.1111/j.2153-3490.1957.tb01849.x.

Rivero-Calle, Sara, Anand Gnanadesikan, Carlos E. Del Castillo, William M. Balch, and Seth D. Guikema. 2015. "Multidecadal Increase in North Atlantic Coccolithophores and the Potential Role of Rising CO2." *Science* 350 (6267): 1533–37. https://doi.org/10.1126/science.aaa8026.

Rogelj, Joeri, Piers M. Forster, Elmar Kriegler, Christopher J. Smith, and Roland Séférian. 2019. "Estimating and Tracking the Remaining Carbon Budget for Stringent Climate Targets." *Nature* 571 (7765): 335–42. https://doi.org/10.1038/s41586-019-1368-z.

Schwartz, Stephen E. 2019. BNL Environmental Science and Technologies Department Seminar: How Long Does Anthropogenic CO2 Stay in the Atmosphere? Brookhaven National Laboratory. https://www.bnl.gov/event.php?q=14749.

Schwartz, Stephen E. 2021. "Observation Based Budget and Lifetime of Excess Atmospheric Carbon Dioxide." *Atmospheric Chemistry and Physics Discussions*, ahead of print, November 29. https://doi.org/10.5194/acp-2021-924.

Schwartz, Stephen E., Quan Hua, David E. Andrews, et al. 2024. "Discussion: Presentation of Atmospheric $^{14}CO_2$ Data." *Radiocarbon* 66 (2): 386–99. https://doi.org/10.1017/RDC.2024.27.

Spencer, Roy. 2023. "ENSO Impact on the Declining CO2 Sink Rate." *J Mari Scie Res Ocean* 6 (4): 163–70. https://doi.org/10.33140/jmsro.06.04.03.

Spencer, Roy W., and John R. Christy. 2024. "Effective Climate Sensitivity Distributions from a 1D Model of Global Ocean and Land Temperature Trends, 1970–2021." *Theoretical and Applied Climatology* 155 (1): 299–308. https://doi.org/10.1007/s00704-023-04634-7.

Stallinga, Peter. 2023. "Residence Time vs. Adjustment Time of Carbon Dioxide in the Atmosphere." *Entropy* 25 (2): 384. https://doi.org/10.3390/e25020384.

Stokes, Nick. 2015. "Why Is Cumulative CO2 Airborne Fraction Nearly Constant?." *Moyhu*, November 28. https://moyhu.blogspot.com/2015/11/why-is-cumulative-co2-airborne-fraction.html.

Strassmann, Kuno M., and Fortunat Joos. 2018. "The Bern Simple Climate Model (BernSCM) v1.0: An Extensible and Fully Documented Open-Source Re-Implementation of the Bern Reduced-Form Model for Global Carbon Cycle–Climate Simulations." *Geoscientific Model Development* 11 (5): 1887–908. https://doi.org/10.5194/gmd-11-1887-2018.

Tahvonen, Olli, Hans von Storch, and Jin-Song von Storch. 1994. "Economic Efficiency of CO2 Reduction Programs." *Climate Reserach* 4 (August): 127–41. https://doi.org/10.3354/cr004127.

Taylor, Charles A., and Wolfram Schlenker. 2025. "Environmental Drivers of Agricultural Productivity Growth: CO2 Fertilization of US Field Crops." *National Bureau of Economic Research*, no. w29320. https://doi.org/10.3386/w29320.

Trenberth, Kevin E., and Lesley Smith. 2005. "The Mass of the Atmosphere: A Constraint on Global Analyses." Journal of Climate. *Journal of Climate* 18 (6): 864–75. https://doi.org/10.1175/JCLI-3299.1.

Uddin, Shihab, Markus Löw, Shahnaj Parvin, et al. 2018. "Elevated [CO2] Mitigates the Effect of Surface Drought by Stimulating Root Growth to Access Sub-Soil Water." *PLOS ONE* 13 (6): e0198928. https://doi.org/10.1371/journal.pone.0198928.

Wang, Jianliang, Lianyong Feng, Xu Tang, Yongmei Bentley, and Mikael Höök. 2017. "The Implications of Fossil Fuel Supply Constraints on Climate Change Projections: A Supply-Side Analysis." *Futures* 86 (February): 58–72. https://doi.org/10.1016/j.futures.2016.04.007.

Wijngaarden, W. A. van, and W. Happer. 2020. "Dependence of Earth's Thermal Radiation on Five Most Abundant Greenhouse Gases." arXiv:2006.03098. Preprint, arXiv, June 4. https://doi.org/10.48550/arXiv.2006.03098.

Zhu, Zaichun, Shilong Piao, Ranga B. Myneni, et al. 2016. "Greening of the Earth and Its Drivers." *Nature Climate Change* 6 (8): 791–95. https://doi.org/10.1038/nclimate3004.